\documentclass[12pt,preprint]{aastex}

\usepackage{graphics,epsfig}
\usepackage{natbib}

\def\spose#1{\hbox to 0pt{#1\hss}}
\def\lta{\mathrel{\spose{\lower 3pt\hbox{$\mathchar"218$}}
     \raise 2.0pt\hbox{$\mathchar"13C$}}}
\def\gta{\mathrel{\spose{\lower 3pt\hbox{$\mathchar"218$}}
     \raise 2.0pt\hbox{$\mathchar"13E$}}}

\def\figure#1#2 {\par{\narrower\noindent {\bf Fig. #1}
   \hskip 2mm #2\par}\bigskip\noindent}
\def\table#1#2 {\par{\narrower\noindent {\bf Tab. #1}
   \hskip 2mm #2\par}\bigskip\noindent}

\shorttitle{Habitability around Main-Sequence Stars}

\shortauthors{Cuntz et al.}

\begin{document}


\title{Habitability of Super-Earth Planets around \\
Main-Sequence Stars including Red Giant Branch Evolution: \\
Models based on the Integrated System Approach}


\vspace{2.5cm}
\author{M. Cuntz$^1$, W. von Bloh$^2$, K.-P. Schr\"oder$^3$, C. Bounama$^2$, S. Franck$^{2, \ast}$}
\vspace{2.5cm}
\affil{$^1$Department of Physics, University of Texas at Arlington, Box 19059, \\
       Arlington, TX 76019, USA}
\vspace{0.5cm}
\affil{$^2$Potsdam Institute for Climate Impact Research, 14412 Potsdam, Germany}
\vspace{0.5cm}
\affil{$^3$Department of Astronomy, University of Guanajuato, 36000 Guanajuato, GTO, Mexico}
\vspace{0.5cm}
\email{cuntz@uta.edu, bloh@pik-potsdam.de, kps@astro.ugto.mx, bounama@pik-potsdam.de}
\vspace{0.5cm}
\affil{$^\ast${\rm deceased}}


\clearpage


\begin{abstract}
In a previous study published in {\it Astrobiology}, 
we focused on the evolution of habitability of a 10~$M_\oplus$
super-Earth planet orbiting a star akin to the Sun.  This study was based on a
concept of planetary habitability in accordance to the integrated system approach
that describes the photosynthetic biomass production taking into account a variety
of climatological, biogeochemical, and geodynamical processes.  In the present
study, we pursue a significant augmentation of our previous work by considering
stars with zero-age main sequence masses between 0.5 and 2.0~$M_\odot$ with
special emphasis on models of 0.8, 0.9, 1.2 and 1.5~$M_\odot$.
Our models of habitability consider again geodynamical processes during the
main-sequence stage of these stars as well as during their red giant branch
evolution.  Pertaining to the different types of stars, we identify so-called
photosynthesis-sustaining habitable zones (pHZ) determined by the limits of
biological productivity on the planetary surface.  We obtain various sets of
solutions consistent with the principal possibility of life.  Considering that
stars of relatively high masses depart from the main-sequence much earlier than
low-mass stars, it is found that the biospheric life-span of super-Earth
planets of stars with masses above approximately 1.5~$M_{\odot}$ is always
limited by the increase in stellar luminosity.  However, for stars with masses
below 0.9~$M_{\odot}$, the life-span of super-Earths is solely determined by
the geodynamic time-scale.  For central star masses between 0.9 and
1.5~$M_{\odot}$, the possibility of life in the framework of our models
depends on the relative continental area of the super-Earth planet.
\end{abstract}


\keywords{
extrasolar planets, geodynamics, habitable zone, planetary climate,
stellar evolution and super-Earths.
}

\clearpage


\clearpage

\section{Introduction}

A central topic of contemporaneous astrobiology is the study of habitability
around different types of stars, particularly main-sequence stars.
This is motivated by the fact that stars spent most of their lifetimes on the
main-sequence \citep[e.g.,][]{maed88}.  Furthermore, it is noteworthy that the
distribution of stellar spectral types is strongly tilted toward low-mass stars,
i.e., K and M-type stars, as implied by detailed determinations of the initial
mass function in the solar neighborhood and beyond \citep[e.g.,][]{krou02,chab03}.
For the present study we want to consider stars of large abundance in the
galactic disk, within a reasonably large mass range, and which all have the
potential of being central objects to habitable planetary systems.  Thus, it
is the aim of the present paper to significantly augment our previous study that
dealt with the evolution of circumstellar habitability for stars akin to the Sun
($1~M_\odot$), see \cite{bloh09}, by considering stars with masses between 0.5
and 2.0~$M_\odot$.  Moreover, we will adopt a wider grid of planetary masses,
ecompassing 1 to 10 Earth masses ($M_\oplus$).

\cite{bloh09} in Paper~I presented a detailed thermal evolution model for a
10 Earth-mass planet in the environment of a star like the Sun while also
adopting an up-to-date solar evolutionary model \citep{schr08}.  The planetary
model of habitability has been based on the integrated system approach, which
describes the photosynthetic biomass production taking into account a variety of
climatological, biogeochemical, and geodynamical processes.  This allowed us to
identify a so-called photosynthesis-sustaining habitable zone (pHZ) determined by
the limits of biological productivity on the planetary surface.  The model also
considered the principle possibility of habitability during stellar evolution
along the Red Giant Branch (RGB).  It was found that the solar pHZ increases in
width over time and moves outward, as expected.  For example, for ages of 11.0,
11.5, 12.0, and 12.1 Gyr, the pHZ is found to extend from 1.41 to 2.60, 1.58
to 2.60, 4.03 to 6.03, and 6.35 to 9.35 AU, respectively.

The approach of evaluating habitability based on assessing the pHZ is alternative
to the study of the climatological habitable zone used by \cite{kast93} and others.
Note that concerning Earth-mass planets, a detailed study of
geodynamic habitability based on the pHZ was previously presented by \cite{fran00b}.
They found that Earth will be rendered uninhabitable after 6.5~Gyr as a result
of plate tectonics, notably the growth of the continental area (enhanced
loss of atmospheric CO$_2$ by the increased weathering surface) and
the dwindling spreading rate (diminishing CO$_2$ output from the
solid Earth).  This work already considered systems of different
types of main-sequence stars; however, it was based on an earlier
version of stellar evolution models given by \cite{scha92}.  Moreover,
it also did not include stellar post--main-sequence evolution.

The study by \cite{fran00b} implies that there is no merit in investigating
the future habitability of Earth during long-term stellar evolution, as
in the framework of pHZ models, the lifetime of habitability is limited by
terrestrial geodynamic processes.  However, this situation is expected
to be significantly different for super-Earth planets due to inherent
differences compared to Earth-mass planets \citep[e.g.,][]{vale07}.
A further motivation for this type of work stems from the ongoing discovery
of super-Earths in the solar neighborhood with the Gliese 876 \citep{rive05}
and Gliese 581 \citep{udry07,vogt10} systems as prime examples. 
In the following, we describe aspects of stellar evolutionary processes
pertaining to our study.  Next, we discuss the definition of the
photosynthesis-sustaining habitable zone, including the relevant geodynamic
assumptions.  Thereafter, we present our results and discussion.  Finally,
we convey our summary and conclusions.

\section{Stellar Evolution Computations}

\subsection{Methods}

As in Paper~I, we base the selected evolution models on the 
well-tested Eggleton code, which allows us to follow the changes of 
stellar properties through the stages of the main-sequence, the 
red giant branch, and beyond.  Our computations have been made with 
an advanced version of the Eggleton code, which considers updated opacities
and an improved equation of state as described by \cite{pols95,pols98}. 
For the abundance of heavy elements, which decisively affect the opacities,
we use the near-solar value of $Z=0.02$.  This choice is an appropriate
representation of present-day samples of stars in the thin galactic disk, 
noting that these stars show a relatively narrow distribution ($Z=0.01$
to 0.03) regarding heavy element abundances about this value.  Besides
other desirable characteristics, the adopted evolution code uses a 
self-adapting mesh and has a treatment of ``overshooting" that has 
thoroughly been tested.  Its two parameters, i.e., the mixing length and the
overshoot length, have been calibrated by utilizing giant and supergiant
stars in well-studied, eclipsing binary systems \citep{schr97}. 

The evolution code, as used by us, also considers a detailed description of 
the stellar mass loss, including its impact on the evolution during the stages
of giant star evolution following \cite{schr05}.  The treatment of the
stellar mass loss rate as function of the governing stellar parameters
has successfully been tested on globular clusters as well as for a set
of well studied stars \citep{schr07}.  Thus, the attained models of
stellar evolution are expected to provide an accurate description of the
time-dependent behavior of stellar luminosity (see Fig.~1), as well
as other stellar quantities, from the main-sequence to the RGB stage, for the
entire range of masses considered in our study.  Also note that as a
consequence of the steadily increasing mass loss, particularly on the
upper RGB of various stars including the Sun, the orbital distance $R$
of any putative planet increases as $R \propto M_*^{-1}$ with $M_*$ as
stellar mass, owing to the conservation of the orbital angular momentum
of the planet.

\subsection{Mass range of interest}

In Table~1, we present some characteristics of stellar evolution
models for the mass range of 0.5 to 2 $M_{\odot}$. For comparison, 
we note that the solar age, according to our models, is given as
4.58 Gyr, see \cite{schr08} (for a metallicity of $Z$ = 0.018), and
the Sun's  present-day effective temperature is identified as
$T_{\rm eff} = 5774$~K.  Masses below 0.5 $M_{\odot}$, i.e., most
M-type stars, have been omitted from our study. Their stellar evolution
can be neglected, luminosity and stellar temperature remain unchanged.

Moreover, stars of relatively large masses, i.e., $M > 2~M_\odot$
are also considered poor candidates for supporting life because the
stellar lifetimes are almost certainly too short to permit the onset
of biology.  In Table~1, we list information on the target stars
taken into consideration.  Particularly, we convey the time for each
stellar model at which the initial (i.e., zero-age main-sequence) luminosity
$L_{\rm ZAMS}$ has increased by 50\%, and when $L_{\rm ZAMS}$ has
doubled.  Stars above $2~M_\odot$ reach the stage at which they
leave the main-sequence phase after less than a billion years.  If we set a
relatively low age limit of, say, 300 million years as a requirement
for the development of complex life forms during the phase of slowly
increasing stellar luminosity, we would limit the mass range of
potential host stars of life-bearing planets to a maximum of
about $1.3~M_\odot$.

In Table~1, we also list the stellar effective temperatures,
which are a steep monotonic function of the stellar mass.  Note
that due to the appearance of increased photospheric UV fluxes,
relatively high effective temperatures by themselves represent an
adverse factor for the general possibility of advanced life forms
\citep[e.g.,][]{cock99,cunt10}, which is particularly relevant for
stars of spectral type mid-F and earlier, although the biological
impact of stellar UV environments is typically significantly altered
by the attenuation of planetary atmospheres.  This consideration is
a further motivation to disregard stars with masses beyond 1.5 or
$2.0~M_\odot$ (i.e., early F stars) in the context of exobiology.
Towards the low mass limit at about $0.5~M_\odot$, there are
virtually no stellar evolutionary changes for these stars while
being on the main-sequence owing to the current age of the Universe.
However, regarding M-type stars, adverse influences on the origin
and development of sophisticated life forms also exist, particularly
associated with strong UV, EUV and X-ray flaring
\citep[e.g.,][]{hawl03,scal07,cunt10,segu10}.

Close to about one solar mass, we find an interesting result: Stars of, e.g., 
$0.9~M_\odot$, still spend many billion years in increasing their luminosity
from 1.5 times to 2 times of their zero-age main-sequence luminosity. In that phase,
which is sufficiently stable and long-lived to provide the development of life forms,
the initially significantly cooler effective temperature of such a star 
is closing in on that of the early Sun. This change of quality and quantity 
of the host star radiation could ``unfreeze'' a planet, which was initially 
outside the habitable zone, and still leave it with enough time to develop 
complex life forms. When reaching about 1.5 times their zero-age main-sequence
luminosity, stars of 0.9 solar masses have reached an age of about 8 billion 
years (see Fig.~2), much like the older stars of the thin galactic disk. 
Hence, several such cases should be found in present-day stellar
samples of the solar neighbourhood. We also note that these same stars 
continue to evolve quite slowly. It will take them several further 
billions of years to leave the main-sequence and to advance on the RGB.

\subsection{A closer look at the mass range of 0.8 to 1.5~$M_\odot$}

As previously pointed out, the most promising types of stars for hosting 
life-bearing planets regarding our grid of models are those of 0.8, 0.9,
1.0 (Sun), 1.2 and $1.5~M_\odot$.  All these stars have a lot in common:
they all evolve into red giants (RGB phase) after central hydrogen burning
has ceased.  The expansion of the outer layers of a RGB star is driven by
two related processes: the contraction of the inner helium core, which has
lost its stabilising energy production, and the growing energy production
by the hydrogen burning shell around it.  This layer right above the
contracting core experiences a steady density increase, which drives up
the resulting stellar luminosity.  The now much higher energy output of
the hydrogen burning shell as well as the much higher density of the core
region lead to a very expanded outer stellar structure, associated with
a significant increase in stellar luminosity and radius (see Fig.~2).
Additionally, the stellar structure is shaped by the onset of a cool wind,
associated with significant mass loss \citep[e.g.,][]{dupr87}.

For our models of 1.0, 1.2 and 1.5 $M_\odot$, the appearance of a helium
flash marks the end of the RGB phase, as the stars settle into a 
new equilibrium with a much less compact core (due to its new source of 
energy) and a, consequently, much less expanded outer structure.  However,
the post-RGB stellar evolution does not offer any promising  aspects for
the origin of life, since the relatively stable phase of central helium
burning lasts only a few hundred million years.  Furthermore, any
ancient life forms on previously habitable worlds will most likely be
destroyed, since in the extreme RGB phases of those stars amount to
200-300 times its initial size and several thousand times its initial
luminosity.  This entails a complete nullification of previous zones of
circumstellar habitability established during stellar main-sequence
evolution.

However, despite such remarkable homogeneity concerning their main physical 
processes, stars in the range of 0.8 to 1.5 solar masses differ largely in 
their lifetimes; see Fig.~2 and Table~1 for details.  On the upper RGB,
the much longer times spent in that phase by low-mass stars have an
interesting consequence: their mass loss experienced in this phase amounts
to more than $0.4~M_\odot$.  This has dramatic implications for any
existing planets at his stages noting that their orbital distance $R$
increases as $R \propto M_*^{-1}$.  This may imply that an at that
time habitable planet may permanently leave the stellar habitable zone or,
conversely, an inhabitable planet may enter the stellar habitable zone,
and becomes ``unfrozen"; see, e.g., \cite{lope05} and \cite{bloh09} for
previous results for stars akin to the Sun.

In summary, it is the stellar main-sequence phase that is most relevant for the 
evolution of life on planets at a suitable distance. The slower the 
star increases its luminosity, the longer a planet stays within the slowly
expanding habitable zone.  Thus, all stars between 0.9 and $1.2~M_\odot$ 
masses are potential host stars for life from the beginning of stellar evolution.
Stars between 0.8 and 0.9 solar masses show an even slower change in stellar
effective temperature and luminosity; in fact, they reach solar luminosity
after 8 billion years.  Detailed studies for the evolution of habitability,
based on the definition of the climatological habitable zones (see below),
during stellar lifetimes were given by \cite{unde03} and others. 
\cite{jone05,jone06} applied this framework to known exoplanetary systems
to evaluate the possibility of habitable ``Earths".  Stellar evolutionary
aspects are also relevant to the future of life on Earth, as discussed by, e.g.,
\cite{sack93} and \cite{schr08}, although the impact of geodynamic processes
appears to be dominant for setting the time limit of terrestrial habitability
\citep{fran00b}.

\section{Habitability of Super-Earth Planets}

\subsection{Photosynthesis-sustaining habitable zone (pHZ)}

Customary simulations about the role of stellar evolution on circumstellar
habitability typically invoke the framework of the climatic habitable zone
\citep{kast93,sels07}.  In this type of models, the zone of habitability
at a given evolutionary time for a star with luminosity $L$ and effective
temperature $T_{\rm eff}$ can be calculated following \cite{unde03} as
\begin{equation}
R_{\mathrm{in}} = \left(\frac{L}{L_\odot\cdot S_{\mathrm{in}}(T_{\rm eff})}\right)^{\frac{1}{2}}~,~~R_{\mathrm{out}}
                = \left(\frac{L}{L_\odot\cdot S_{\mathrm{out}}(T_{\rm eff})}\right)^{\frac{1}{2}}
\label{hz_jones}
\end{equation}
with $S_{\mathrm{in}}(T_{\rm eff})$ and $S_{\mathrm{out}}(T_{\rm eff})$ described
as second order polynomials.

It is the purpose of this paper to expand the focus of Paper~I, which is solely
focused on the Sun.  Paper~I is based on the concept that the habitability of
super-Earth planets is evaluated via a modified Earth-system model that describes the
evolution of the temperature and atmospheric CO$_2$ concentration.  On Earth, the
carbonate--silicate cycle is the crucial element for a long-term homeostasis
under increasing solar luminosity.  On geological time-scales, the deeper parts
of the Earth are considerable sinks and sources of carbon.

Our numerical model was previously applied to the super-Earths Gl~581c and Gl~581d
\citep{bloh07}, the putative super-Earth Gl~581g \citep{bloh11} as well as fictitious
Earth-mass planets for 47~UMa and 55~Cnc \citep{cunt03,fran03,bloh03}, among numerous
other studies.  This model couples the stellar luminosity $L$, the silicate--rock
weathering rate $F_{\mathrm{wr}}$ and the global energy balance to obtain
estimates of the partial pressure of atmospheric carbon dioxide
$P_{\mathrm{CO}_2}$, the mean global surface
temperature $T_{\mathrm{surf}}$, and the biological productivity $\Pi$ as
a function of time $t$ (Fig.~\ref{boxmodel}).  The main point is the
persistent balance between the CO$_2$ sink in the atmosphere--ocean system and
the metamorphic (plate tectonic) sources.  This is expressed through the
dimensionless quantities  
\begin{equation}
f_{\mathrm{wr}}(t) \cdot f_A(t) = f_{\mathrm{sr}}(t),
\label{gfr}
\end{equation}
where $f_{\mathrm{wr}}(t) \equiv F_{\mathrm{wr}}(t)/F_{\mathrm{wr},0}$ is the 
weathering rate, $f_A(t) \equiv A_c(t)/A_{c,0}$ is the continental area, and
$f_{\mathrm{sr}}(t) \equiv S(t)/S_0$ is the areal spreading rate, which are all
normalized by their present values of Earth.  
Eq.~(\ref{gfr}) can be rearranged by introducing the geophysical forcing ratio
GFR \citep{volk87} as
\begin{equation}
f_{\mathrm{wr}}(T_{\mathrm{surf}},P_{\mathrm{CO}_2})=\frac{f_{\mathrm{sr}}}{f_A} \ =: \ \mathrm{GFR}(t) \ .
\label{gfr2}
\end{equation}
Here we assume that the weathering rate only depends on the global surface temperature
and the atmospheric CO$_2$ concentration.
For the investigation of a super-Earth under external forcing,
we adopt a model planet with a prescribed continental area.  The fraction of
continental area with respect to the total planetary surface $f_A$ is varied between
$0.1$ and $0.9$.

The connection between the stellar parameters and the planetary climate can be
obtained by using a radiation balance equation \citep{will98}
\begin{equation}
\frac{L}{4\pi R^2} [1- a (T_{\mathrm{surf}}, P_{\mathrm{CO}_2})]
 = 4I_R (T_{\mathrm{surf}}, P_{\mathrm{CO}_2}),
\label{L}
\end{equation}
where $a$ denotes the planetary albedo, $I_R$ the outgoing infrared flux, and $R$ the distance
from the central star.   The Eqs.~(\ref{gfr2}) and (\ref{L}) constitute a set of two coupled
equations with two unknowns, $T_{\mathrm{surf}}$ and $P_{\mathrm{CO}_2}$, if the parameterization
of the weathering rate, the luminosity, the distance to the central star and the geophysical forcing
ratio are specified.  Therefore, a numerical solution can be attained in a straightforward manner.

Resulting from the described model the domain of distances $R$ for a super-Earth planet from its
respective main-sequence star for which a photosynthesis-active biosphere is productive
($\Pi > 0$) can be calculated.  It is defined as the photosynthetis-sustaining habitable zone
given as
\begin{equation}
\mathrm{pHZ}(t):=\{ R | \Pi (P_{\mathrm{CO}_2}(R,t),T_{\mathrm{surf}}(R,t))>0\}.
\end{equation}
In our model, biological productivity is considered to be solely a function of
the surface temperature and the CO$_2$ partial pressure in the atmosphere.
Our parameterization yields zero productivity for $T_{\mathrm{surf}} \leq 0^{\circ}$C or $T_{\mathrm{surf}}
\geq 100^{\circ}$C or $P_{\mathrm{CO}_2}\leq 10^{-5}$ bar \citep{fran00a}.
The inner and outer boundaries of the pHZ do not depend on
the detailed parameterization of the biological productivity within the temperature
and pressure tolerance window.

The maximum life span of a photosynthetic-active biosphere $t_{\mathrm{max}}$ is the
point in time when a planet at an optimum distance from its central star
$R_{\mathrm{opt}}$ finally leaves the pHZ
\begin{equation}
t_\mathrm{max}=\max_t |\mathrm{pHZ}(t)|>0. \label{eq:tmax}
\end{equation}
Note that the evaluation of $\mathrm{pHZ}(t)$ and $t_\mathrm{max}$ will be essential
to our models of circumstellar habitability.

\subsection{Comments on the thermal evolution model}

Parameterized convection models are the simplest models for the investigation of the thermal
evolution of terrestrial planets and satellites. They have successfully been applied to the 
evolution of Mercury, Venus, Earth, Mars, and the Moon \citep{stev83,slee00}.
\cite{fran95} studied the thermal and volatile history of Earth and Venus in the
framework of comparative planetology. The internal structure of massive terrestrial planets
with one to ten Earth masses has been investigated by \cite{vale06} to
obtain scaling laws for total radius, mantle thickness, core size, and average density as 
a function of mass. Similar scaling laws were found for different compositions. We will
use these scaling laws for mass-dependent properties of super-Earths and also the mass-independent
material properties previously given by \cite{fran95}.

The thermal history and future of a super-Earth planet has to be determined to
calculate the spreading rate for solving key Eq.~(\ref{gfr}).
A parameterized model of whole mantle convection including the volatile exchange
between the mantle and surface reservoirs \citep{fran95,fran98} is applied.
The key equations used in our present study are in accord with our previous work
focused on Gl~581c, Gl~581d, and Gl~581g; see \cite{bloh07,bloh11} for details.
A key element is the computation of the areal spreading rate $S$; note that $S$
is a function of the average mantle temperature $T_m$, the 
surface temperature $T_{\mathrm{surf}}$, the heat flow from the mantle $q_m$, and the
area of ocean basins $A_0$ \citep{turc02}.  It is given as
\begin{equation}  S = \frac{q_m^2 \pi
\kappa A_0}{4 k^2 (T_m - T_\mathrm{surf})^2}\,, 
\end{equation}
where $\kappa$ is the thermal diffusivity and $k$ the thermal conductivity.
To calculate the spreading rate, the thermal evolution of the mantle has be to computed:
\begin{equation} {4 \over 3} \pi \rho c (R_m^3-R_c^3) \frac{dT_m}{dt} = -4 \pi
R_m^2 q_m + {4 \over 3} \pi E(t) (R_m^3-R_c^3), \label{therm} \end{equation}
where $\rho$ is the density, $c$ is the specific heat at constant pressure,
$E$ is the energy production rate by
decay of radiogenic heat sources in the mantle per unit volume, and $R_m$ and $R_c$ are the
outer and inner radii of the mantle, respectively.
To calculate the thermal evolution for a planet with several Earth masses the planetary
parameters have to be adjusted.  Therefore, we assume
\begin{equation}
\frac{R_p}{R_{\oplus}}= \left(\frac{M}{M_{\oplus}}\right)^{0.27}
\end{equation} and
where $R_p$ is the planetary radius, see \cite{vale06}.

The total radius, mantle thickness, core size and average density are all functions
of mass, with the subscript $\oplus$ denoting Earth values.
The exponent of $0.27$ has been obtained for super-Earths.  The values
of $R_p$, $R_m$, $R_c$, as well as the other planetary properties are scaled
accordingly.  Table~2 gives a summary of these size parameters for the planetary
models between 1 and $10~M_\oplus$; see the studies by \cite{bloh07,bloh11}
for additional information including parameters for the mass-independent
quantities.  According to \cite{vale09}, we assume that a more
massive planet is likely to convect in a plate tectonic regime similar to Earth.
Thus, the more massive the planet, the higher the Rayleigh number that controls
convection, the thinner the top boundary layer (lithosphere), and the higher the
convective velocities.

We recognize that there is an ongoing debate about the onset and significance of
plate tectonics on super-Earth planets.  On one hand, \cite{stei04} and \cite{onei07}
pointed out that there might be in an episodic or stagnant lid regime, considering
that increasing the planetary radius acts to decrease the ratio of driving to
resisting stresses, an effect that appears to be robust when increases in planetary
gravity are included.  Other thermodynamic processes that peak at different planet
masses \citep[e.g.,][]{noac11} operate as well.  On the other hand, \cite{vale07} and
\cite{vale09} argued that planet tectonics of super-Earths should be considered
inevitable.  Their models show that as the planetary mass increases, the shear stress
to overcome resistence to plate thickness increases while the plate thickness decreases,
thereby enhancing plate weakness.  These effects contribute favorably to the subduction
of the lithosphere (like on Earth), a crucial component of plate tectonics.  Noting
that in the framework of our model, a plate-tectonic-driven carbon cycle is considered
necessary and essential for carbon-based life, we will assume the existence of
plate tectonics on super-Earths for our models in the following.

\section{Results and Discussion}

\subsection{Habitability based on the integrated system approach}

We calculated the photosynthesis-sustaining habitable zones for a $10~M_\oplus$
super-Earth with relative continental areas between $r=0.1$ and 0.9 orbiting central
stars in the mass range between 0.8 and 1.5~$M_\odot$.  The results are shown in
Fig.~\ref{hzint}.  The width of the pHZ during the main-sequence evolution is found
to be approximately constant for all stellar models taken into account.
For stellar mass $M<0.9~M_\odot$ the pHZ ceases to exist before the
main-sequence evolution ends, while for higher stellar masses the width of the pHZ
increases significantly and moves outwardly in time.  The time when this processes
occurs critically depends on the stellar mass, as expected.

``Water worlds", i.e., planets mostly covered by oceans are favored in the
facilitation of habitability during this stage as previously obtained in the
context of long-term evolution for hypothetical Earth-mass and super-Earth planets
in various star-planet systems \citep{fran03,bloh03,bloh07,bloh09}.  In principle,
the favored existence of water worlds instead of land worlds around stars during
their phase of red giant branch evolution constitutes a possible extension of the
outer edge of circumstellar habitability.  The reason for the favorism of water worlds
under those circumstances is that planets with a considerable continental area have
relatively high weathering rates that provide the main sink of atmospheric CO$_2$.
Therefore, this type of planets are unable to build up CO$_2$-rich atmospheres that
prevent the planet from freezing, which in turn is thwarting the possibility of
photosynthesis-based life.

The maximum life spans $t_{\mathrm{max}}$ (Eq.~\ref{eq:tmax}) of
super-Earth planets have been calculated for a grid of planetary masses between
1 and $10~M_\oplus$ with increments of $1~M_\oplus$ and relative continental areas $r$
between 0.1 and 0.9.  Figure~(\ref{lifespan}) depicts $t_\mathrm{max}$ as a function
of the planetary mass.  It is found that $t_{\mathrm{max}}$ increases with planetary
mass and, furthermore, decreases for increasing $r$ values.  In addition,
we depict the time-spans as a function of the stellar mass when the central star reaches
a luminosity of $L=1.5~L_{\rm ZAMS}$ and $L=2~L_{\rm ZAMS}$ with $L_{\rm ZAMS}$ as the
initial (i.e., zero-age main-sequence) luminosity at time $t=0$. 

Moreover, in Fig.~\ref{domain} the critical central stellar mass is depicted dependent
on the planetary mass up to which the life span of the biosphere is solely determined
by $t_{\mathrm{max}}$ (colored shaded areas) instead of the increase in stellar
luminosity (white area).  It is found that the biospheric life span of super-Earth
planets for central stars with masses above about $1.5 M_{\odot}$ is always limited
by the increase in stellar luminosity.  For central star masses below $0.9 M_{\odot}$
it is solely determined by $t_{\mathrm{max}}$, i.e., the maximal geodynamic life span
of the biosphere can always be realised irrespectively of the stellar nuclear evolution.
For central star masses between 0.9 and $1.5 M_{\odot}$ the situation depends on the
relative continental area $r$.

Main-sequence stars with masses between 0.5 and $0.9~M_{\odot}$ belong to late-type
G, K, and early M stars.  The luminosity of such stars remains almost
constant during their entire nuclear evolution on the main-sequence. Therefore, planets
within the habitable zone would remain habitable for a very long period of time if
the geodynamic evolution of the planet is not taken into account.  The planetary
evolution is driven by the long-term cooling of the planetary interior.  Therefore,
diminished internal forcing will affect the planetary habitability of the biospheres.
In the framework of the adopted models it is found that the maximum life span
of super-Earth planets around low mass stars depends almost entirely on the properties
of the planet.  It is known, however, that other factors are also expected to
potentially impact or limit the habitability in the environments of late K and M stars,
encompassing effects associated with, e.g., tidal locking of the planet and
stellar flares \citep[e.g.,][]{lamm07,tart07,sels07,lamm09,segu10,hell11}.

\section{Summary and Conclusions}

We investigated the habitability of super-Earth planets in the environments
of main-sequence stars of masses between 0.5 and and 2.0~$M_\odot$.  This work
complements a previous study that has been focused on star akin to the Sun
($M = 1 M_\odot$), see \cite{bloh09}.  Both studies employ the concept of
planetary habitability based on the integrated system approach that describes
the photosynthetic biomass production taking into account a variety of
climatological, biogeochemical, and geodynamical processes.
Among the various assumptions entertained as part of our study, we
consider that super-Earth planets are expected to develop plate tectonics,
an assumption critical to our models, although it is currently under
scrutiny.  Pertaining to the
different types of stars, we identify so-called photosynthesis-sustaining
habitable zones (pHZ) determined by the limits of biological productivity on
the planetary surface.  We obtain various sets of solutions consistent with
the principal possibility of life.  Our models are based on advanced version
of the Eggleton stellar evolution code that considers updated opacities and 
an updated equation of state as described by \cite{pols95,pols98}. 
Additionally, an improved description of mass loss is implemented following
the work by \cite{schr05}.

Considering that high-mass stars depart from the main-sequence much
faster than low-mass stars, it was found that the biospheric life-span of
super-Earth planets of stars with masses above approximately 1.5~$M_{\odot}$
is always limited by the increase in stellar luminosity dictated by stellar
nuclear evolution.  However, for stars with masses below 0.9~$M_{\odot}$,
the life-span of super-Earths it is solely determined by the geodynamic
time-scale.  Clearly, the luminosity of stars with spectral type mid-K or
later has not deviated much from their ZAMS luminosities owing to the
limited age of the Universe.  For central star masses between 0.9 and
1.5~$M_{\odot}$, the possibility of life in the framework of our models
depends on the relative continental area of the super-Earth planets.

A crucial finding of our study is that planets of large geological age are expected
to be water worlds, i.e., planets with a relatively small continental area.  The
reason why land worlds are considered to be unfavorable is that planets with a
considerable continental area have relatively high weathering rates that provide
the main sink of atmospheric CO$_2$.  Therefore, this type of planets are unable
to build up CO$_2$-rich atmospheres that prevent the planet from freezing or
allowing for the possibility of photosynthesis-based life.  Our study is
also both expanding and superseding previous work, including the study by
\cite{fran00b}.  This work solely considered Earth-mass planets instead of
also including super-Earth planets.  Additionally, it considered a 
linear growth model for the continental growth and has refrained from taking
into account post-main sequence evolution.  Although the evolutionary
time-scales for stars on the RGB are short compared to the time spent on
the main-sequence, habitable planets around red giants should be
considered a realistic possibility, as previously pointed out by
\cite{lope05}, \cite{bloh09}, and others.  However, for central stars with
higher masses than the Sun, a more rapid evolution will occur that will also
place the significant temporal and spatial constraints on planetary habitability
when the central stars have reached the RGB.

\pagebreak

\newpage


\begin{figure*}
\centering
\epsfig{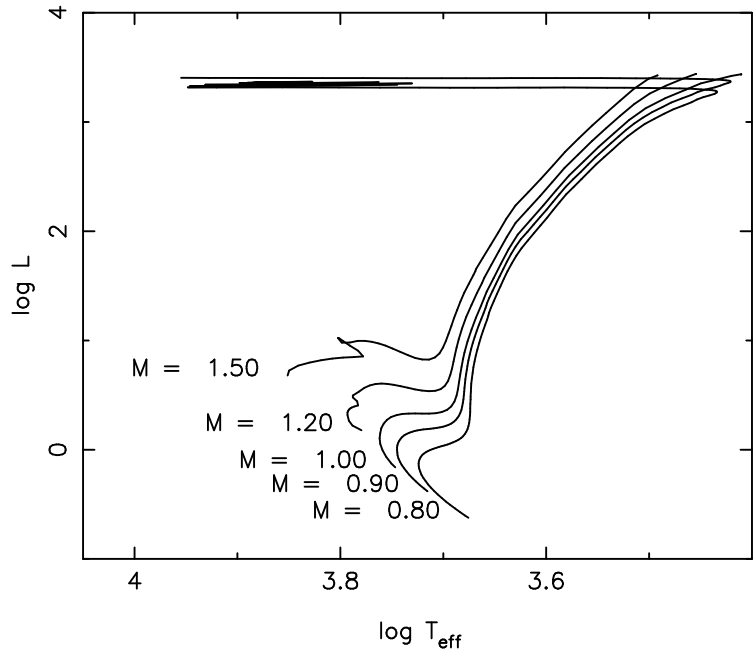}
\caption{
Hertzsprung-Russell diagram ($\log{L}$ versus $\log{T_{\rm eff}}$)
of stellar evolution tracks with initial masses of 0.8, 0.9, 1.0 (Sun),
1.2 and 1.5~$M_\odot$, reaching from the ZAMS to the tip of the RGB.
In the first two cases, the stars lose so much mass on the RGB 
that they do not ignite He-burning but directly evolve into white dwarfs.
The AGB branches of the more massive stars have been omitted for the sake
of clarity.
\label{HRDmodel}
}
\end{figure*}

\begin{figure*}
\centering
\begin{tabular}{c}
\epsfig{file=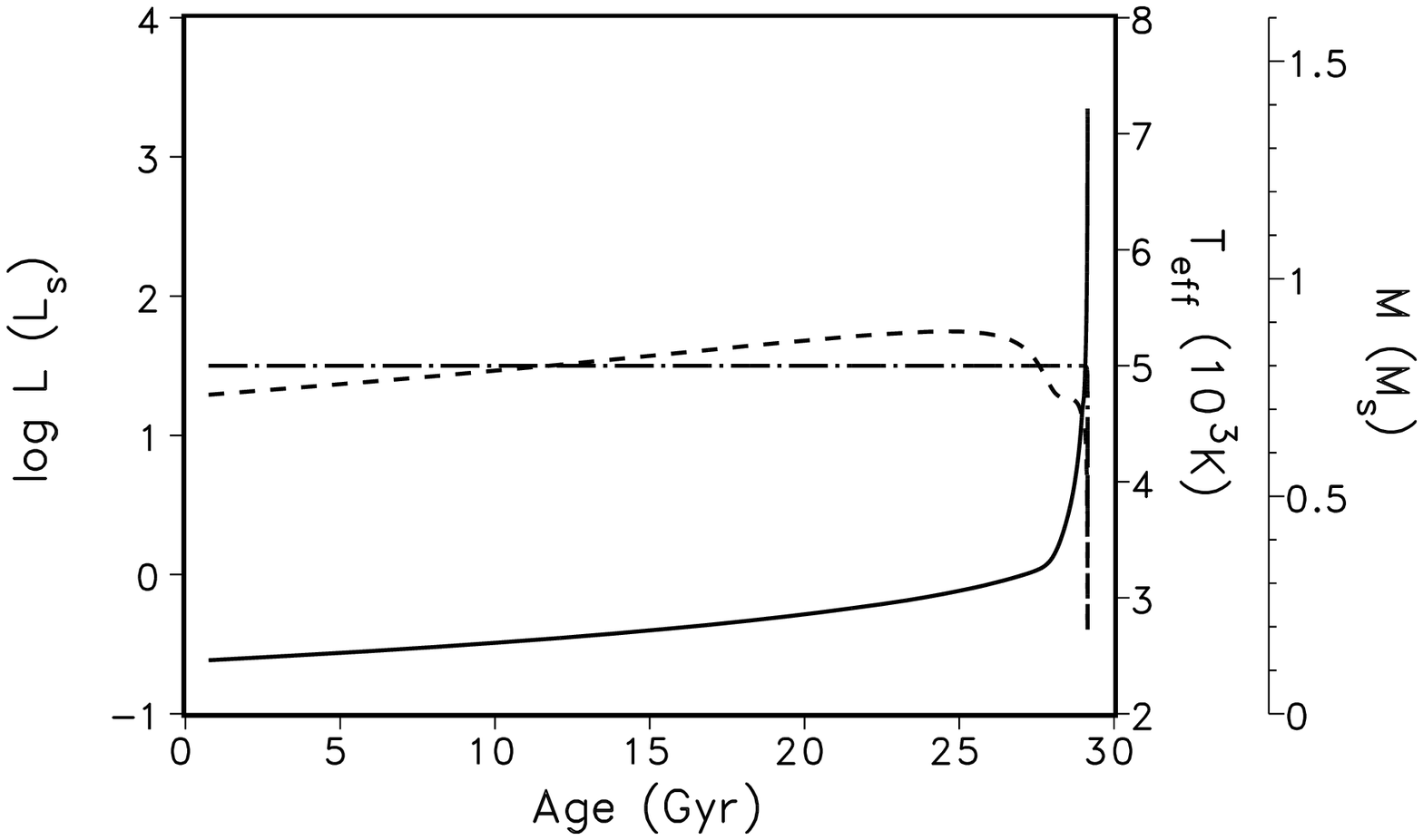,width=0.45\linewidth} \\
\epsfig{file=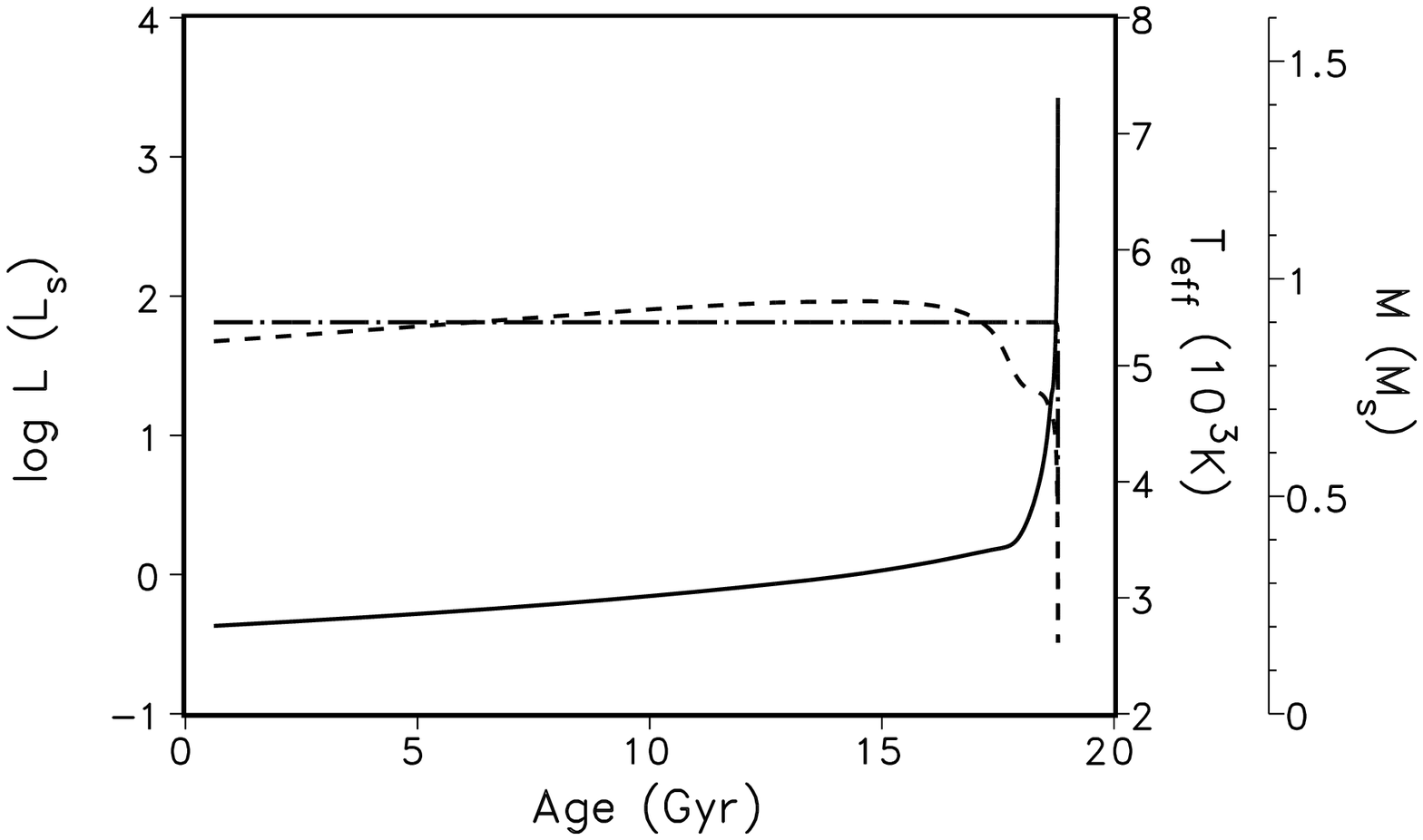,width=0.45\linewidth} \\
\epsfig{file=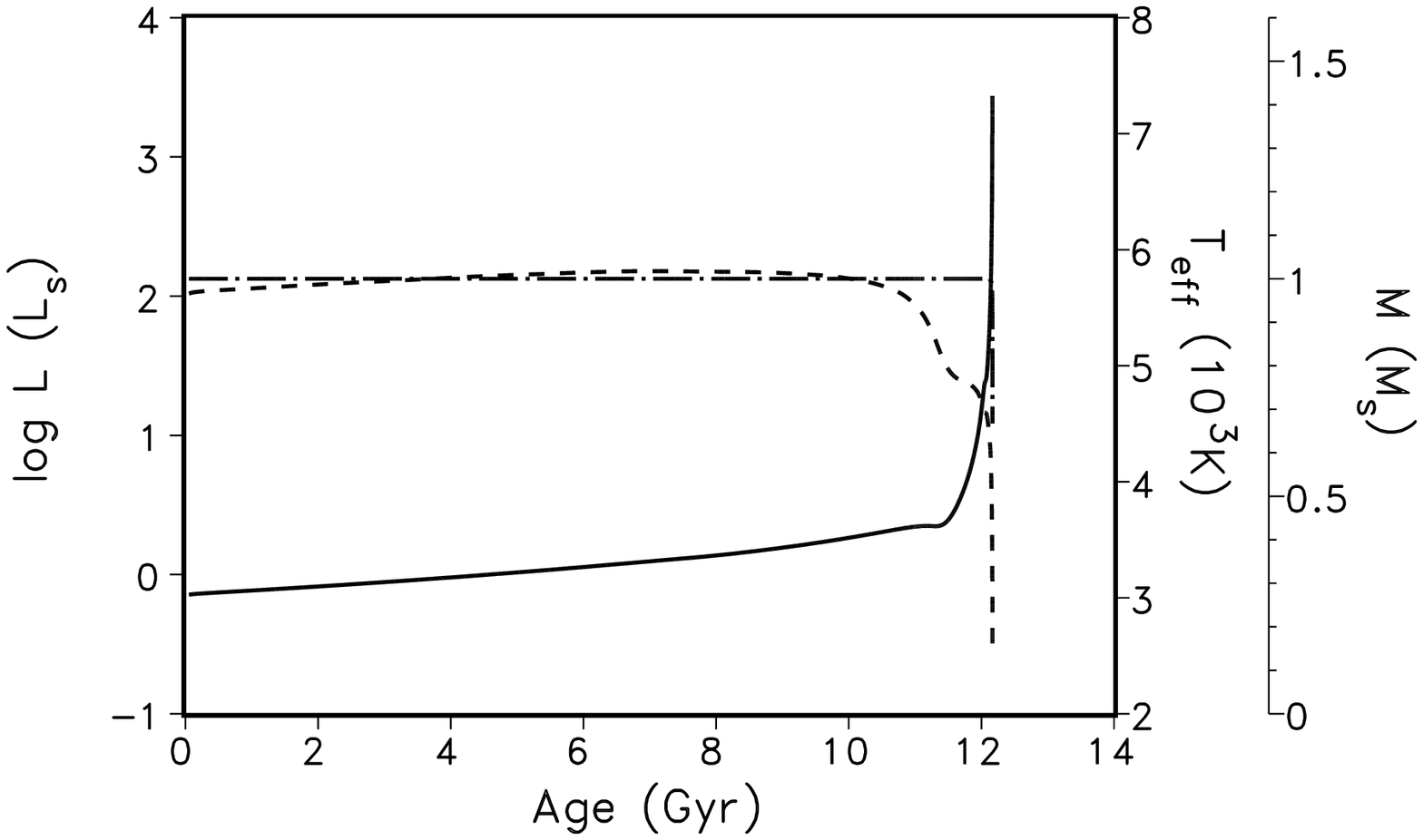,width=0.45\linewidth} \\
\epsfig{file=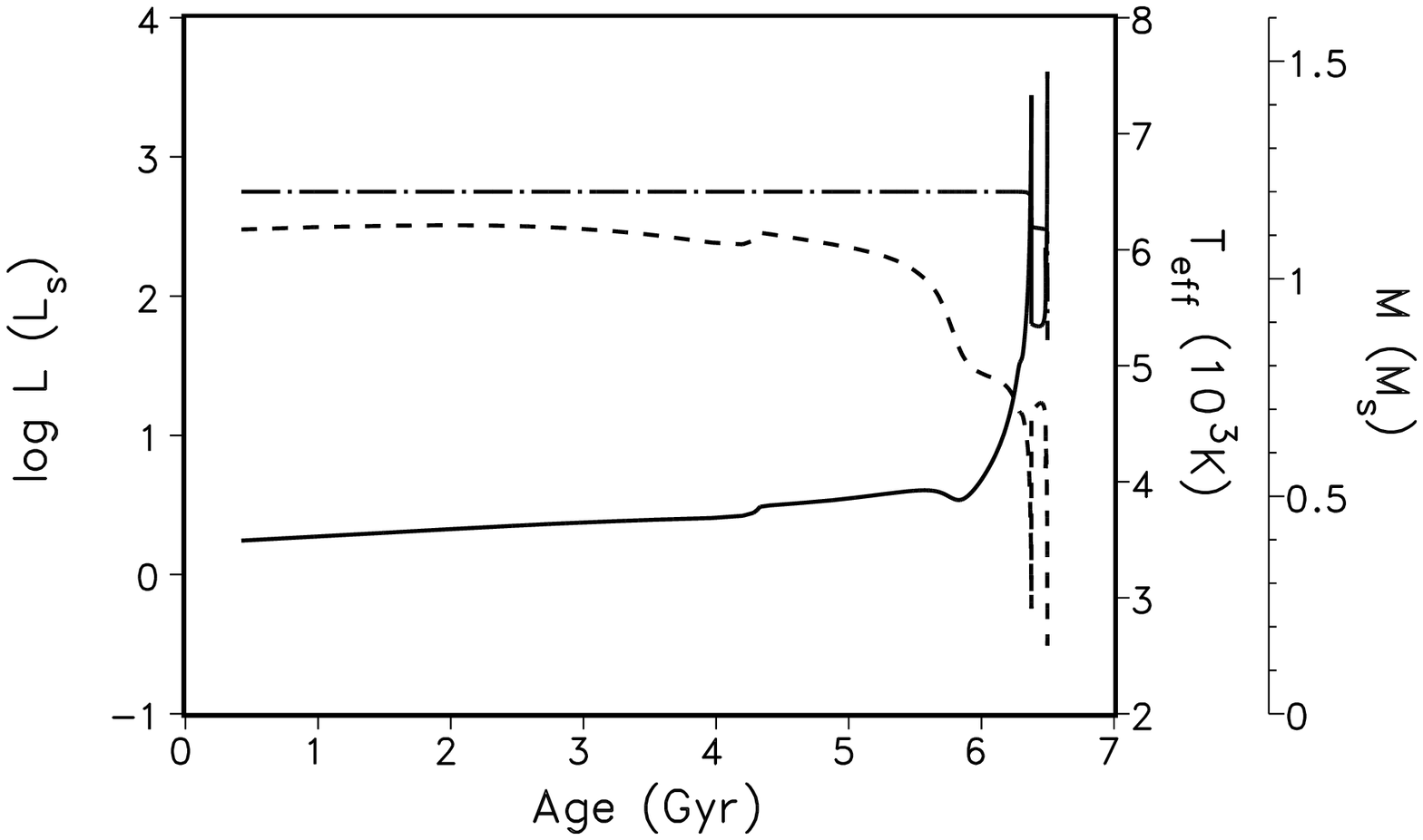,width=0.45\linewidth} \\
\epsfig{file=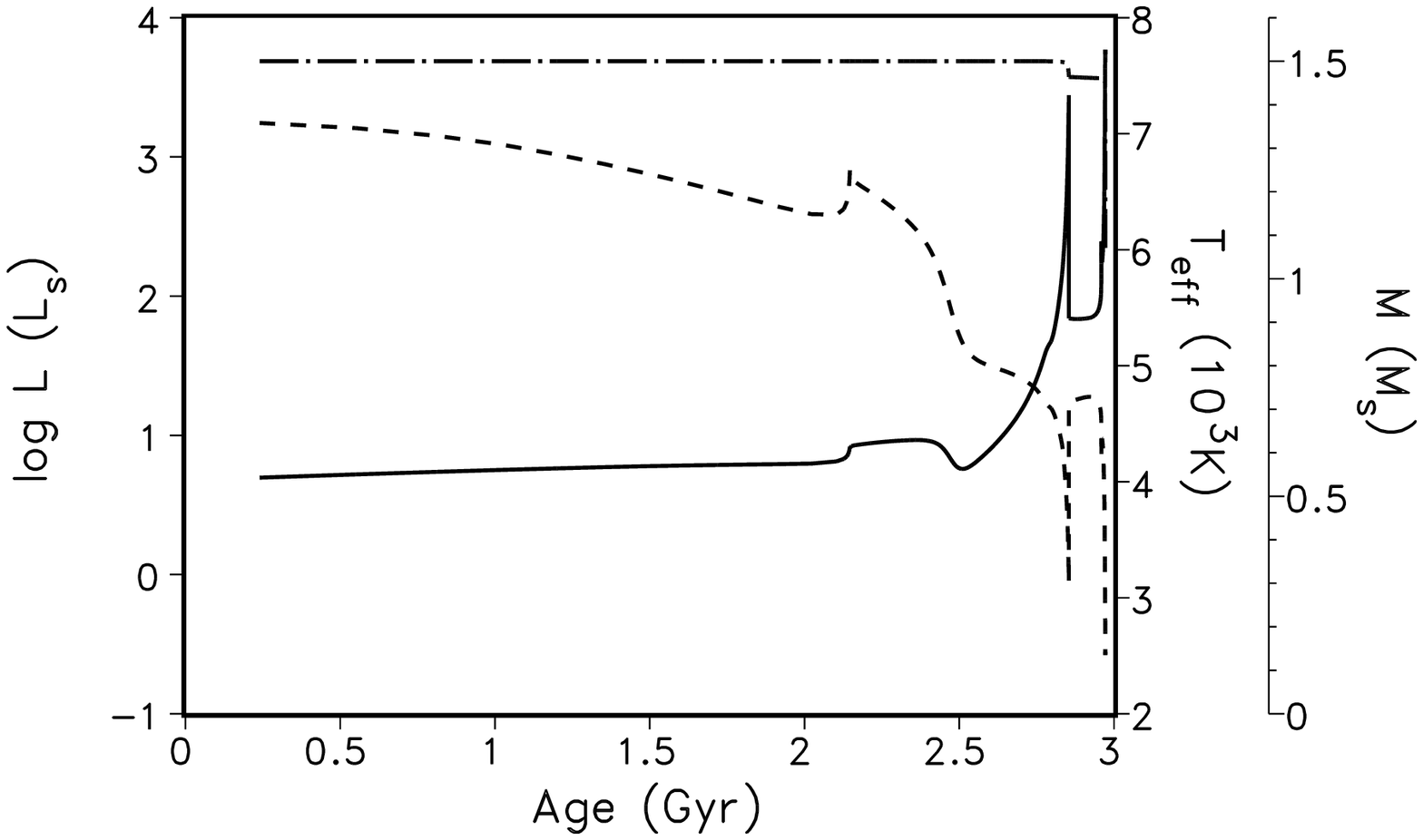,width=0.45\linewidth} \\
\end{tabular}
\caption{Stellar evolution models following \cite{schr08},
depicting the luminosity (solid line), the effective temperature
(dashed line), and the mass (dash-dotted line).  Note the
vast differences in the $x$-axes within the figure panel.
}
\label{lumin}
\end{figure*}

\begin{figure*}
\centering
\epsfig{file=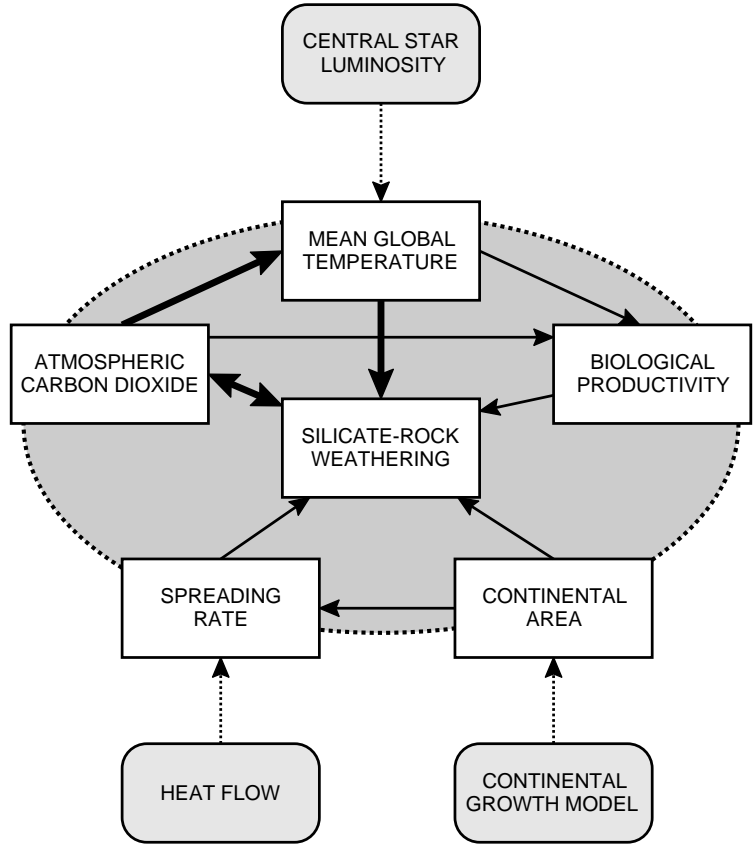,width=0.60\linewidth}
\caption{
Box model of the integrated system approach adopted in
our model.  The arrows indicate the different types of forcing,
which are the main feedback loop for stabilizing the climate
(thick solid arrows), the feedback loop within the system
(thin solid arrows), and the external and internal forcings
(dashed arrows).
\label{boxmodel}
}
\end{figure*}

\begin{figure*}
\centering
\epsfig{file=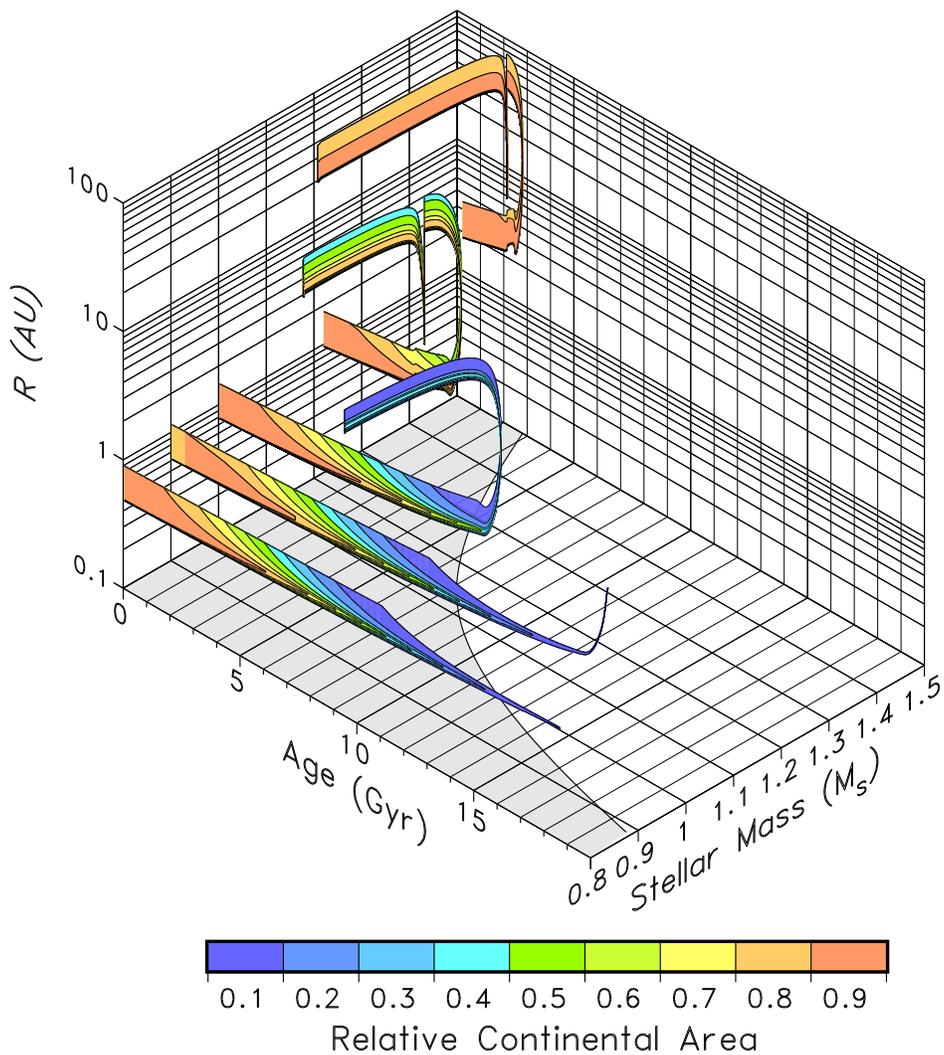,width=0.75\linewidth}
\caption{
The pHZ for a 10 Earth-mass planet orbiting an $M=0.8,0.9,1.0,1.2,1.5~M_\odot$
star also considering post-main sequence evolution. The gray shaded area at the 
bottom denotes the time period of the stellar main-sequence evolution.
Note the ``bending-over" of the color-coded areas for stars with masses of
1.0~$M_\odot$ and above due to the drastic increases in the stellar luminosities.
\label{hzint}
}
\end{figure*}

\begin{figure*}
\centering
\epsfig{file=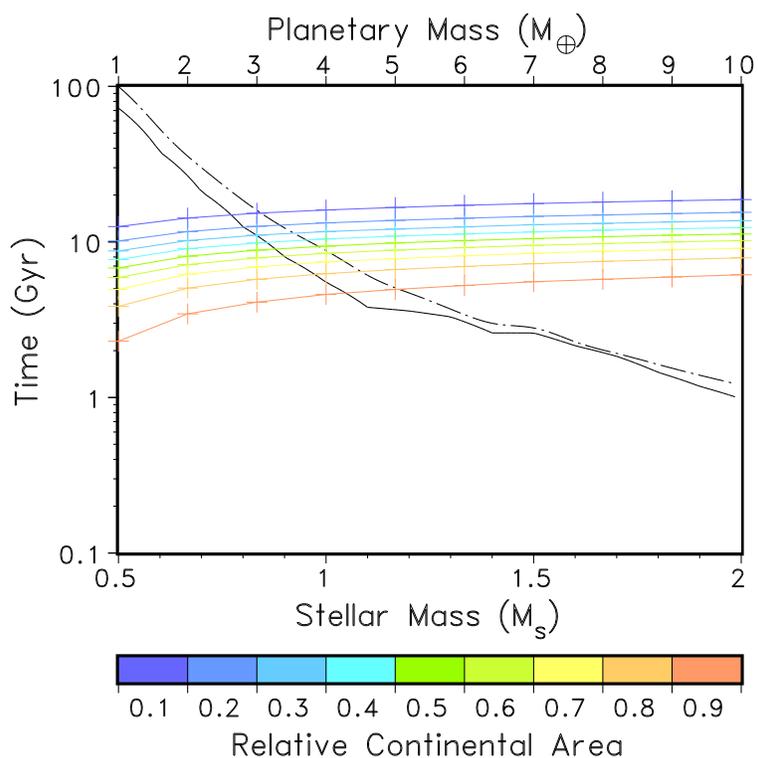,width=0.60\linewidth}
\caption{
The maximum life span of the biosphere $t_{\mathrm{max}}$
in dependence on the planetary mass $M$ (colored solid lines).
The graphs are color coded by the assumed portion of the planetary surface
covered by continents $r$.  The solid black line denotes the time when
the star reaches a luminosity of $L=1.5~L_{\rm ZAMS}$, while the dashed line
denotes the time for $L=2.0~L_{\rm ZAMS}$.
}
\label{lifespan}
\end{figure*}

\begin{figure*}
\centering
\epsfig{file=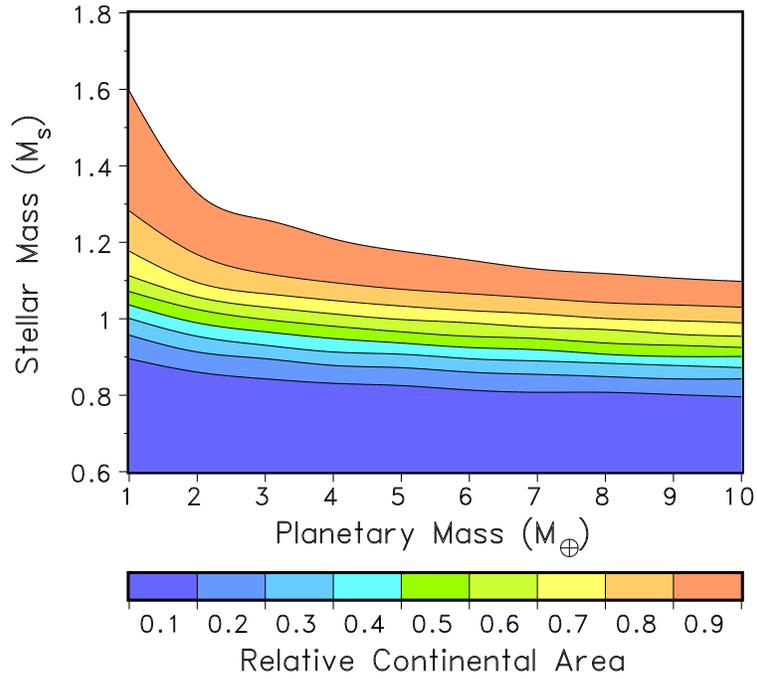,width=0.60\linewidth}
\caption{
The critical mass of the central star as a function of the
planetary mass up to which the life span of the biosphere is solely
determined by the maximum geodynamic life span (color-coded area)
instead of the time of the stellar nuclear evolution, i.e., when
the star reaches a luminosity of $L=2.0~L_{\rm ZAMS}$
(white area).  Different colors correlate to different relative
continental areas of the model planet.
}
\label{domain}
\end{figure*}


\clearpage

\begin{deluxetable}{lcccccc}
\tablecaption{Stellar Parameters\label{tabstars}}
\tablewidth{0pt}
\tablehead{
   $M$         &  $\log(L_{\rm ZAMS})/L_\odot)$  & $T_{\rm eff}$  & Age($1.5 \times L_{\rm ZAMS})$ &  $T_{\rm eff}$  & Age($2 \times L_{\rm ZAMS})$  & $T_{\rm eff}$   \\
   ($M_\odot$) &  ...  & (K)  & (Gyr)  &  (K)  & (Gyr)  & (K)
}
\startdata
  2.0   &     1.204    &        9160    &       0.98   &     7390    &     1.20  &     7100  \\
  1.9   &     1.113    &        8790    &       1.18   &     6975    &     1.39  &     6885  \\
  1.8   &     1.017    &        8400    &       1.45   &     6455    &     1.63  &     6380  \\
  1.7   &     0.914    &        7990    &       1.84   &     6130    &     1.92  &     6260  \\
  1.6   &     0.802    &        7555    &       2.15   &     6080    &     2.28  &     6340  \\
  1.5   &     0.679    &        7100    &       2.6    &     6000    &     2.8   &     6300  \\
  1.4   &     0.543    &        6710    &       2.6    &     6220    &     3.0   &     6180  \\
  1.3   &     0.391    &        6420    &       3.3    &     6145    &     3.7   &     6165  \\
  1.2   &     0.221    &        6140    &       3.6    &     6110    &     4.7   &     6070  \\
  1.1   &     0.015    &        5820    &       3.8    &     5990    &     6.1   &     5960  \\
  1.0   &   $-0.163$   &        5580    &       5.5    &     5750    &     8.8   &     5770  \\
  0.9   &   $-0.383$   &        5180    &       8.0    &     5430    &    12.3   &     5530  \\
  0.8   &   $-0.625$   &        4735    &      12.5    &     5015    &    18.5   &     5185  \\
  0.7   &   $-0.897$   &        4260    &      21.0    &     4540    &    30.0   &     4720  \\
  0.6   &   $-1.170$   &        3930    &      37.8    &     4110    &    53.1   &     4260  \\
  0.5   &   $-1.430$   &        3750    &      72.5    &     3865    &    99.8   &     3960  \\
\enddata
\end{deluxetable}

\clearpage

\begin{deluxetable}{lcccccl}
\tablecaption{Stellar and Planetary Parameters\label{tabplan}}
\tablewidth{0pt}
\tablehead{
Parameter & \multicolumn{4}{c}{Value}   & Unit & Description \\
 ...      & $1~M_\oplus$ & $2~M_\oplus$ & $5~M_\oplus$ & $10~M_\oplus$ & ...  & ...
}
\startdata
$g$         &      1.00  &      1.38  &     2.10  &       2.88  & $g_\oplus$  & gravitational acceleration  \\
$R_p$       &   6378     &   7691     &  9849     &  11,876     &   m         & planetary radius            \\
$R_c$       &   3471     &   4185     &  5360     &    6463     &   m         & inner radius of the mantle  \\
$R_m$       &   6271     &   7562     &  9684     &  11,677     &   m         & outer radius of the mantle  \\
\enddata
\end{deluxetable}

\end{document}